\documentclass[11pt,a4paper]{article}
\usepackage{jheppub}
\usepackage[T1]{fontenc}

\allowdisplaybreaks

\newcommand{\be}{\begin{equation}}
\newcommand{\ee}{\end{equation}}
\newcommand{\bea}{\begin{eqnarray}}
\newcommand{\eea}{\end{eqnarray}}

\title{Proton decay suppression in a supersymmetric SO(10) model}
\author{Xiaojia Li}
\author{and Da-Xin Zhang}
\affiliation{School of Physics
and State Key Laboratory of Nuclear Physics and Technology,\\
Peking University, Beijing 100871, China}
\emailAdd{shakalee@pku.edu.cn}
\emailAdd{dxzhang@pku.edu.cn}

\abstract{
We propose a mechanism for sufficient suppression of dimension-5 operators for
proton decay  in a supersymmetric SO(10) model.
This mechanism is analogue to the double seesaw mechanism in studying neutrino masses.
Only an intermediate VEV instead of
an intermediate scale is required so that gauge coupling unification is maintained.
The VEV is generated by introducing an anomalous U(1) symmetry whose breaking is at higher scale.
The proton decay amplitudes are suppressed by this VEV over the GUT scale.
We use \textbf{45+54} in breaking GUT symmetry.
\textbf{120} is included so that fermion sector is fully realistic.
Assuming a minimal fine-tuning in the Higgs doublet sector,
$\textrm{tan}\beta$ of order one is predicted. }
\keywords{SO(10), proton decay, supersymmetry}

\arxivnumber{1409.5233}
\begin{document}
\maketitle
\flushbottom
\pagenumbering{arabic}

\section{Introduction} \label{introduction}
Grand Unified Theory (GUT)\cite{gut1,gut2} is one of
the most attractive candidates for the physics beyond the Standard Model (SM).
The Supersymmetric (SUSY) GUT models based on SO(10)\cite{clark1982,aulakh1983}
are especially interesting for several reasons.
Firstly, each generation of fermion superfields are unified in
a single \textbf{16}-plet spinor representation which contains the right-handed neutrino,
so that sub-eV neutrino masses can be generated naturally by the seesaw mechanism\cite{seesawi1,seesawi2,seesawi3,seesawi4,seesawi5,seesawii1,seesawii2,seesawii3,seesawii4}.
Secondly, in the renormalizable versions of SUSY SO(10) models\cite{rp1,rp2,rp3},
R-parity is conserved automatically which
eliminates the most dangerous dimension-4 operators of proton decay.

The elegant running behaviors of the coupling constants in MSSM
strongly suggest that the unification scale should be taken at $2\times 10^{16}$GeV \cite{Einhorn,Marciano,MSSM1,MSSM2,MSSM3,MSSM4}
which we call the GUT scale $M_G$.
Not only the three coupling constants are unified at $M_G$, but also the masses of the gauge superfields G(SO(10))/G(SM) are taken at the same scale. It has been recognized recently\cite{dlz} that instead of an intermediate seesaw scale,
in SUSY SO(10) models with several pairs of \textbf{126} $+\overline{\textbf{126}}$,
only an intermediate vacuum expectation value (VEV) of the SM singlet in one $\overline{\textbf{126}}$
is needed which couples with the matter superfields.
Consequently, the spectra of this kind of  models do not contain particles at
intermediate scale so that gauge coupling unification is maintained,
meanwhile the seesaw mechanism still works.
This mechanism is further incorporated in models aiming at sufficiently suppressing proton decay\cite{dlz2},
where the seesaw VEV is related to the VEV of an SO(10) singlet which breaks an extra global U(1) symmetry.
Proton decay amplitudes are found to be suppressed in \cite{dlz2} by a factor $\frac{M_I}{M_G}$,
where $M_I\sim 10^{14}$GeV is the seesaw VEV which is much smaller than the GUT scale $M_G\sim 10^{16}$GeV.
This suppression of proton decay is archived by the enhancements of the effective triplet masses
through an inversely analogue to the mass texture in the seesaw mechanism, or a lever mechanism.

In the present work we will extend the observation made by the previous study that the seesaw VEV might be
related to the suppression of proton decay in other models. Instead of using \textbf{210} to break SO(10),
we will use \textbf{45}+\textbf{54}.
The global U(1) will be replaced by an anomalous U(1) whose breaking is generated
by an SO(10) singlet through the Green-Shwarz mechanism\cite{green1,green2}.
The enhancements of the effective triplet masses responsible for proton decay are through
an inversely analogue to the mass texture in the double-seesaw mechanism\cite{double,double2}.
%, or a double-lever mechanism.
We will not improve on either the running behavior of the SO(10) gauge coupling above the GUT scale
or on the minimal fine-tuning for the weak doublets which is implicitly assumed.

In the next section, we will give a simple overview on proton decay suppression.
Then, we will propose in Section \ref{model}
a renormalizable model and show its consistency with high energy supersymmetry.
Proton decay suppression mechanism in this model is shown in Section \ref{proton}.
The discussion on the weak doublets of the MSSM and the prediction of small $\textrm{tan}\beta$
are followed in Section \ref{DT}.
We will summarize in Section \ref{Sum}.
%==========================================================================

\section{General consideration on proton decay suppression}
Consider a simplified model with two pairs of color triplets-anti-triplets,
with only one pair of them couple with fermions. The mass term for the triplets can be written as
$(\varphi_{\overline{T}})_i(M)_{ij}(\varphi_T)_j$,
where $i$, $j$ run from 1 to 2. We need to rotate to the mass eigenstates in order to calculate proton decay amplitudes. Two $2\times2$ unitary matrices $U$ and $V$ are introduced as
\begin{equation}%\label{}
M'_{ij}=U_{ik}M_{kl}{V}^{\dagger}_{lj},
\end{equation}
where $M'=\textrm{diag}(M_1, M_2).$
The mass eigenstates $\varphi'_{\overline{T}}$s and $\varphi'_{T}$s are
\begin{equation}%\label{}
(\varphi'_{\overline{T}})_i=(\varphi_{\overline{T}})_j{U}^{\dagger}_{ji},\  \ (\varphi'_{T})_i={V}_{ij}(\varphi_{T})_j.
\end{equation}
Then the dimension-5 operators mediated by the color triplet higgsinos are proportional to \cite{Nath}
\begin{equation}\label{m-1}
\sum_i{V}^{\dagger}_{1i}\frac{1}{M'_{ii}}{U}_{i1}=({V}^{\dagger}\cdot M'^{-1}\cdot{U})_{11}=[({U}^{\dagger}\cdot M'\cdot{V})^{-1}]_{11}=(M^{-1})_{11}.
\end{equation}
The inverse of  $(M^{-1})_{11}$ is called the effective triplet mass which mimics the role of the
color triplet higgsino in the simplest models with only one pair of color triplet-anti-triplet.

Eq.(\ref{m-1}) is easy to be generalized to models with more pairs of color triplets-anti-triplets.
The proton decay amplitudes are proportional to sums of specific elements  in  the inverse of the triplet mass matrix.
Algebraically, these matrix elements in the inverse mass matrix  can be written as
 \begin{equation}\label{m-1ij}
(M^{-1})_{ij}=(-1)^{i+j}\frac{\textrm{Det}(M^*_{ji})}{\textrm{Det}(M)},
\end{equation}
where $M^*_{ij}$ represents $M$ with the $i$th row and the $j$th column eliminated,
whose determinant is called as the algebraic complement,
and $i$ and $j$ are the labels of those color triplets-anti-triplets which can couple with the fermions.

There are two possible ways to get small $(M^{-1})_{ij}$'s following (\ref{m-1ij}).
We can construct a mass matrix either with all small algebraic complements for the elements which couple with
fermions, or with a large determinant of the entire mass matrix.
In the previous work \cite{dlz2} the first approach is used where the color triplet mass matrix
can be symbolically expressed as
\begin{equation}\label{m2p2}
M_T=\left(
\begin{array}{cc}
0 &M_{G} \\
M_{G} & M_I
\end{array}
\right).
\end{equation}
Here $M_G$ stands for a GUT scale mass while $M_I$ is the intermediate seesaw VEV.
Only the up-left block couples with matter fields, so it is clear that $M^*_{11}=M_I$ is smaller than $M_G$.

In this work, we are trying to realize the second possibility. The mass matrix for the color triplets is written as
\begin{equation}\label{m3p3}
M_T=\left(
\begin{array}{ccc}
0 &M_{G} &0\\
M_{G} & 0 &M_{G} \\
0 &M_{G} & M_{P}
\end{array}
\right).
\end{equation}
Again, the matter fields couple with the up-left block only.
Here $M_{P}$ represents a mass at a scale higher than the GUT scale, or at the Plank scale.
Then $\textrm{Det}(M_T)\sim M_{P}M_{G}^2$ is enhanced to give large effective triplet masses.
%But the result is slightly different, see Eq.(\ref{trisym1}).

As the texture  in (\ref{m2p2}) is analogue to the neutrino mass matrix in the seesaw mechanism,
the present texture  in (\ref{m3p3}) is analogue to the neutrino mass matrix in the double-seesaw mechanism.
The mass texture either in (\ref{m2p2})  or in (\ref{m3p3}) is sufficient to suppress
proton decay.

\section{The model and SUSY preserving} \label{model}
The particle content of the present model is as follows.
First, it contains three generations of fermion fields
which are embedded into three \textbf{16}-plet ($\psi_{1,2,3}$) superfields as usual.
Second, \textbf{45}+\textbf{54} ($A$, $E$) are introduced to break SO(10).
Third, in order to give satisfied fermion masses and mixing \cite{Bajc},
Higgs in \textbf{120} ($D$) is introduced, which is also needed to couple through \textbf{45}+\textbf{54}
with those in \textbf{10} ($H$) and in \textbf{126}/$\overline{\textbf{126}}$ ($\Delta/\overline{\Delta}$).
Forth, the \textbf{45} is further copied ($A'$) to generate a small VEV for the seesaw mechanism, and to generate the structure (\ref{m3p3}) for suppression of proton decay. Three sets of Higgs are needed with the first two sets contain $H+ \Delta/\overline{\Delta} +D$
while the third set contains $\Delta/\overline{\Delta}$.
An extra U(1) symmetry,
whose breaking is realized by the SO(10) singlets $S_1$ and $S_2$,
is introduced to distinguish these Higgs.
All the fields and their U(1) charges are listed in Table \ref{Qnumbers}.
Note that the different U(1) charges of the first set of Higgs ($H_1...$) and
the third set of Higgs ($\Delta_3...$) also require different fields ($A+E$ and $A'$)
to couple the first two and the last two sets of Higgs.
\begin{table}[t]
\begin{center}
\begin{tabular}{|c|c|c|c|c|c|c|c|c|}
\hline
 &$\psi_{i}$ & $H_1,D_1,\Delta_1/\overline{\Delta}_1$ &  $H_2,D_2,\Delta_2/\overline{\Delta}_2$ &  $\Delta_3/\overline{\Delta}_3$ & $A,E $ & $A'$ & $S_1$ &$S_2$ \\
\hline
U(1) charge &$-\frac{1}{2}$ &1 &-1 &$\frac{1}{2}$ &0 &$\frac{1}{2}$ &-1 &$-\frac{1}{2}$ \\
\hline
\end{tabular}
\caption{SO(10) multiplets and their U(1) charges.} \label{Qnumbers}
\end{center}
\end{table}

Only $H_1,D_1,\overline{\Delta}_1$ couple with matter fields due to the U(1) charges.
The Yukawa sector is given as
\begin{equation}\label{superp1}
W_Y=Y_{10}^{ij}\psi_i\psi_j H_1+Y_{120}^{ij}\psi_i\psi_j D_1+Y_{126}^{ij}\psi_i\psi_j \overline{\Delta}_1,
\end{equation}
which is general enough to fit all fermion masses and mixing \cite{rp2,fits1,fits2,fits3,fits4,Bajc2004,Bajc2006,revise1,revise2,revise3}.

The general renormalizable Higgs superpotential is given by
\begin{eqnarray}  \label{superp2}
W&=&m_H H_1H_2+{m_{\Delta}}_{12}\overline{\Delta}_1\Delta_2
+{m_{\Delta}}_{21}\overline{\Delta}_2\Delta_1+m_D D_1D_2+\frac{1}{2}m_{A}A^2+\frac{1}{2}m_{E}E^2 \nonumber \\
&+&H_1H_2(\lambda_1A +\lambda_2E)-i A(\lambda_{3}\overline{\Delta}_1\Delta_2+\lambda_{4}\overline{\Delta}_2\Delta_1)
+ E(\lambda_{5}{\Delta}_1\Delta_2+\lambda_{6}\overline{\Delta}_1\overline{\Delta}_2)\nonumber \\
&+&D_1 A(\lambda_7H_2+\lambda_{8}\Delta_2+\lambda_{9}\overline{\Delta}_2)
+D_2 A(\lambda_{10}H_1+\lambda_{11}\Delta_1+\lambda_{12}\overline{\Delta}_1)\nonumber \\
&+&D_1D_2(\lambda_{13}A +\lambda_{14}E)+\lambda_{15} E^3+\lambda_{16} AE^2 -i A'(\alpha_{1}\overline{\Delta}_2\Delta_3+\alpha_{2}\overline{\Delta}_3\Delta_2)\nonumber \\
&+&D_2 A'(\alpha_{3}\Delta_3+\alpha_{4}\overline{\Delta}_3)+
\frac{1}{2}S_1(2\beta_{1}\overline{\Delta}_3\Delta_3+\beta_2{A'}^2)+\beta_3 S_2 AA'.
\end{eqnarray}
Labeled by the representations under the $SU(4)_C\times SU(2)_L\times SU(2)_R$ subgroup
of SO(10), the following components get VEVs responsible for the SO(10) symmetry breaking
\begin{eqnarray}\label{allvev}
A_1^{(\prime)}&=&\langle A^{(\prime)}(1,1,3)\rangle, {}~A_2^{(\prime)}=\langle A^{(\prime)}(15,1,1)\rangle,
{}~E=\langle E(1,1,1)\rangle;  \nonumber \\
v_{(1,2,3)}&=&\langle \Delta_{(1,2,3)} (\overline{10},1,3)\rangle,{}~\overline{v}_{(1,2,3)}
=\langle\overline{\Delta}_{(1,2,3)}(10,1,3)\rangle.
\end{eqnarray}

Inserting these VEVs into (\ref{superp2}), we get
\begin{eqnarray}  \label{superp2VEV}
\langle W\rangle &=&{m_{\Delta}}_{12}\overline{v}_{1}v_2 +{m_{\Delta}}_{21}\overline{v}_{2}v_1 +\frac{1}{2}m_A(A_1^2+A_2^2) +\frac{1}{2}m_E E^2+A_0 (\lambda_{3}\overline{v}_1v_2+\lambda_{4}\overline{v}_2v_1)\nonumber\\
&+&\frac{\lambda_{15}}{2\sqrt{15}} E^3
+\lambda_{16}E(\frac{\sqrt{3}}{2\sqrt{5}}A_1^2-\frac{1}{\sqrt{15}}A_2^2)+A'_0 (\alpha_{1}\overline{v}_2v_3+\alpha_{2}\overline{v}_3v_2)\nonumber \\
&+&\frac{1}{2}S_1(2\beta_{1}\overline{v}_3v_3+\beta_2{A'_1}^2+\beta_2{A'_2}^2
)+S_2(\beta_3A_1A'_1+\beta_3A_2A'_2),
\end{eqnarray}
where we have defined
\begin{eqnarray}\label{A0}
A_0  \equiv  \left(-\frac{1}{5}A_1-\frac{3}{5\sqrt{6}}A_2\right), \quad  A'_0 \equiv \left(-\frac{1}{5}A'_1-\frac{3}{5\sqrt{6}}A'_2\right),
\end{eqnarray}
for later convenience.

To preserve SUSY at high energy, the F- and D-flatness conditions are required.
The D-flatness condition requires
\begin{equation}  \label{dterm}
|v_{1}|^2+|v_{2}|^2+|v_{3}|^2=|\overline{v}_{1}|^2+|\overline{v}_{2}|^2+|\overline{v}_{3}|^2,
\end{equation}
which constrains only the sum of $|v|^2$s and $|\overline{v}|^2$s,
so that an intermediate valued VEV of $\overline{v}_1$ can be generated without breaking gauge coupling
unification, if both sides in (\ref{dterm}) are of the order $M^2_{G}$.

The F-flatness conditions
\begin{equation}
\left\{
\frac{\partial }{\partial v_{1}},
\frac{\partial }{\partial v_{2}},
\frac{\partial }{\partial v_{3}},
\frac{\partial }{\partial \overline{v}_{1}},
\frac{\partial }{\partial \overline{v}_{2}},
\frac{\partial }{\partial \overline{v}_{3}},
\frac{\partial }{\partial A' _{1}},
\frac{\partial }{\partial A' _{2}},
\frac{\partial }{\partial S_{1}},
\frac{\partial }{\partial S_{2}},
\frac{\partial }{\partial A_{1}},
\frac{\partial }{\partial A_{2}},
\frac{\partial }{\partial E }
\right\} \langle W\rangle =0,\nonumber
\label{partiale}
\end{equation}
are explicitly
\begin{eqnarray}
0&=&M_{21}\overline{v}_{2},                                       \label{equv1}\\
0&=&M_{12}\overline{v}_{1}+\alpha_2 A'_0\overline{v}_{3} ,                      \label{equv2} \\
0&=&\alpha_1 A'_0 \overline{v}_{2}+\beta_{1}S_1\overline{v}_{3},                  \label{equv3} \\
0&=&M_{12}{v}_{2},                                     \label{equv1bar}\\
0&=&M_{21}{v}_{1}+\alpha_1 A'_0{v}_{3} ,               \label{equv2bar} \\
0&=&\alpha_2 A'_0 {v}_{2}+\beta_{1}S_1{v}_{3},                                  \label{equv3bar} \\
0&=&\beta_2S_1A'_1+\beta_3S_2A_1-\frac{1}{5}(\alpha_{1}\overline{v}_2v_3+\alpha_{2}\overline{v}_3v_2),        \label{equAp1} \\
0&=&\beta_2S_1A'_2+\beta_3S_2A_2-\frac{3}{5\sqrt6}(\alpha_{1}\overline{v}_2v_3+\alpha_{2}\overline{v}_3v_2), \label{equAp2} \\
0&=&\beta_{1}\overline{v}_3v_3+\beta_2{A'_1}^2+\beta_2{A'_2}^2   \label{equS1} \\
0&=&\beta_3A_1A'_1+\beta_3A_2A'_2     \label{equS2}\\
0&=&m_{A}A_1-\frac{1}{5}(\lambda_{3}\overline{v}_1v_2+\lambda_{4}\overline{v}_2v_1)+\frac{\sqrt3}{\sqrt5}\lambda_{16}EA_1+\beta_3A'_1S_2,  \label{equA1} \\
0&=&m_{A}A_2-\frac{3}{5\sqrt6}(\lambda_{3}\overline{v}_1v_2+\lambda_{4}\overline{v}_2v_1)-\frac{2}{\sqrt{15}}\lambda_{16}EA_2+\beta_3A'_2S_2, \label{equA2} \\
0&=&m_E E+\frac{3}{2\sqrt{15}}\lambda_{15}E^2+\lambda_{16}(\frac{\sqrt{3}}{2\sqrt{5}}A_1^2-\frac{1}{\sqrt{15}}A_2^2),
\label{equE}
\end{eqnarray}
where
\begin{eqnarray}\label{M12M21}
M_{12}\equiv  {m_{\Delta}}_{12}+\lambda_3 A_0, \quad M_{21}\equiv  {m_{\Delta}}_{21}+\lambda_4 A_0.
\end{eqnarray}
From (\ref{dterm}) to (\ref{equE}), there are 13 variables and 14 equations in total,
but only 12 of the equations are independent.
One of the VEVs, $S_1$ for example, can be assigned to any scale.

First, (\ref{equv1})-(\ref{equv3}) are linear equations about the $\overline{v}$s, which can be rewritten as
\begin{equation}\label{detvb}
\left(
\begin{array}{ccc}
\overline{v}_1, &\overline{v}_2, &\overline{v}_3
\end{array}
\right)
\left(
\begin{array}{ccc}
0 &M_{12} &0\\
M_{21} & 0 &\alpha_1 A'_0 \\
0 & \alpha_2 A'_0 &\beta_{1}S_1
\end{array}
\right)=0.
\end{equation}
Similarly, equations (\ref{equv1bar})-(\ref{equv3bar}) can be rewritten as
\begin{equation}\label{detv}
\left(
\begin{array}{ccc}
0 &M_{12} &0\\
M_{21} & 0 &\alpha_1 A'_0 \\
0 & \alpha_2 A'_0 &\beta_{1}S_1
\end{array}
\right)
\left(
\begin{array}{c}
{v}_1 \\
{v}_2 \\
{v}_3
\end{array}
\right)=0.
\end{equation}
Both (\ref{detvb}) and (\ref{detv}) require
\begin{equation}\label{det}
\beta_{1}S_1 M_{12} M_{21}=0,
\end{equation}
which corresponds to three different possibilities as following.
If $S_1=0$, a lot of particles cannot get masses through couplings with $S_1$.
Thus this case is simply excluded.
If $M_{12}=0$ is taken, then from (\ref{detvb}-\ref{detv}) it follows that
$\overline{v}_{1}=M_{G}$ and $\overline{v}_{2}=\overline{v}_{3}=0$ which gives too small
neutrino masses and is thus excluded as well.
We thus have the last possibility,
\begin{equation}\label{M21}
M_{21}=0,
\end{equation}
which gives
\begin{equation}\label{vbratio}
v_1\sim M_G, ~v_2=v_3=0, ~~~\frac{\overline{v}_{1}}{\overline{v}_{3}}=-\frac{\alpha_2 A_0'}{M_{12}}, ~\frac{\overline{v}_{2}}{\overline{v}_{3}}=-\frac{\beta_1 S_1}{\alpha_1 A_0'},
\end{equation}
following (\ref{detvb}) and (\ref{detv}). Then, $\overline{v}_{1,2,3}$ are expressed by $v_1$ through (\ref{dterm}).

Furthermore, substituting $v_2=v_3=0$ into (\ref{equAp1})-(\ref{equAp2}),
$A'_1$ and $A'_2$ can be expressed by $A_1$ and $A_2$, respectively,
and (\ref{equS2}) is now identical to (\ref{equS1}).
Equation $M_{21}=0$ in (\ref{M21}) itself gives the dependence of $A_2$ on $A_1$ through (\ref{M12M21}).
Taking $S_1$ as free, the remaining variables are now $A_1, E, S_2, v_1$
with equations (\ref{equS1}), (\ref{equA1}),(\ref{equA2}) and (\ref{equE}) left. Given the parameters in (\ref{superp2}), all the VEVs are now determined. Numerically, $A_1$, $A_2$ and $E$ are taken as GUT scale VEVs  in order to break SO(10) down to MSSM.

According to the analysis in \cite{missingvev},
the extra U(1) symmetry  is naturally related to string theory,
and it is appropriate to take the VEV of breaking this U(1)  at
$$S_1\sim 10^{17}\textrm{GeV}\sim 10M_G.$$
After inserting (\ref{equAp1}-\ref{equAp2}) into  (\ref{equA1}-\ref{equA2}), the last terms will change into
$$\beta_3A'_1S_2 \rightarrow -\frac{\beta_3^2 S_2^2}{\beta_2 S_1}A_1, \quad \beta_3A'_2S_2 \rightarrow -\frac{\beta_3^2 S_2^2}{\beta_2 S_1}A_2.$$
They are naturally at the same scale as other terms, i.e. $M_G^2$, which indicates that
$$S_2\sim \frac{1}{\sqrt{10}}S_1 \sim \sqrt{10}M_G, \quad A'_1 \sim  A'_2 \sim\frac{1}{\sqrt{10}} M_G.$$
Thus we  get from (\ref{vbratio})
\begin{equation}\label{vvev}
\left(
\begin{array}{ccc}
{v}_1, &{v}_2, &{v}_3
\end{array}
\right)= \left(
\begin{array}{ccc}
O(M_G), &0, &0
\end{array}
\right)
\quad
\left(
\begin{array}{ccc}
\overline{v}_1, &\overline{v}_2, &\overline{v}_3
\end{array}
\right)=
O\left(
\begin{array}{ccc}
10^{-2}M_G, &M_G, &10^{-\frac{3}{2}}M_G
\end{array}
\right).
\end{equation}
Now that all the constrains on SUSY preserving have been satisfied,
all the VEVs can be determined and all their scales are known.
The seesaw VEV $\overline{v}_1\sim M_I$ is naturally generated at $10^{-2}M_G$,
which differs from \cite{dlz2} where it introduced a VEV of an SO(10) singlet
which broke a global U(1) symmetry.

Consequence of the large masses of the third set of Higgs and $A'$ given by the VEV $S_1\sim 10M_G$
is that these Higgs are to be integrated out
above the GUT scale $M_G$, so that they do not affect the running behaviors of gauge couplings of MSSM.

\section{Proton decay suppression} \label{proton}
To demonstrate the effectiveness of the present model on solving the proton decay problem,
we need to write down the color triplet mass matrix.
The color triplets are ordered as
\begin{equation}\label{phit}
\varphi_{T}=(H_{1T},D_{1T},D'_{1T},\Delta_{1T}, \overline{\Delta}_{1T}, \overline{\Delta}'_{1T}; H_{2T},D_{2T},D'_{2T},\Delta_{2T}, \overline{\Delta}_{2T}, \overline{\Delta}'_{2T}; \Delta_{3T}, \overline{\Delta}_{3T}, \overline{\Delta}'_{3T}),
\end{equation}
while the color anti-triplets are
\begin{equation}\label{antiphit}
\varphi_{\overline{T}}=(H_{1\overline{T}},D_{1\overline{T}},D'_{1\overline{T}}, \overline{\Delta}_{1\overline{T}}, \Delta_{1\overline{T}}, \Delta'_{1\overline{T}};
H_{2\overline{T}},D_{2\overline{T}},D'_{2\overline{T}}, \overline{\Delta}_{2\overline{T}},
\Delta_{2\overline{T}}, \Delta'_{2\overline{T}};
\overline{\Delta}_{3\overline{T}}, \Delta_{3\overline{T}}, \Delta'_{3\overline{T}}).
\end{equation}
The mass term of the Higgs color triplets is  given by $(\varphi_{\overline{T}})_a(M_T)_{ab}(\varphi_T)_b$,
 with the $15\times 15$ matrix $M_T$ written as
\begin{equation}\label{triplet}
M_T=\left(
\begin{array}{ccc}
0_{(6\times 6)} &B_{12(6\times 6)} &0_{(6\times 3)}\\
B_{21(6\times 6)} & 0_{(6\times 6)} &B_{23(6\times 3)} \\
0_{(3\times 6)} & B_{32(3\times 6)} &B_{33(3\times 3)} \\
\end{array}
\right),
\end{equation}
where
\begin{equation}\label{MT12}
B_{12}=\left(
\begin{array}{cccccc}
B_{H12} &-\frac{i\lambda_{10}}{\sqrt3}A_1 & -\frac{i\sqrt2\lambda_{10}}{3}A_2 & 0 &0 &0 \\
\frac{i\lambda_{7}}{\sqrt3}A_1 & B_{D12} & 0 & -\frac{i\lambda_{8}}{2\sqrt5}A_1 & -\frac{i\lambda_{9}}{2\sqrt5}A_1 & -\frac{i\lambda_{9}}{\sqrt{15}}A_2\\
-\frac{i\sqrt2\lambda_{7}}{3}A_2 & 0  & B'_{D12} &\frac{i\lambda_{8}}{\sqrt{30}} A_2 &-\frac{i\lambda_{9}}{\sqrt{30}} A_2 &-\frac{i\lambda_{9}}{\sqrt{10}} A_1 \\
0 & \frac{i\lambda_{12}}{2\sqrt5}A_1 & -\frac{i\lambda_{12}}{\sqrt{30}} A_2 & {m_{\Delta}}_{12}+\frac{\lambda_3}{5\sqrt6}A_2 & \frac{2\lambda_6}{\sqrt{15}}E & 0\\
0 & \frac{i\lambda_{11}}{2\sqrt5}A_1 & \frac{i\lambda_{11}}{\sqrt{30}} A_2 & \frac{2\lambda_5}{\sqrt{15}}E & {m_{\Delta}}_{21}-\frac{\lambda_4}{5\sqrt6}A_2 & 0\\
0 & \frac{i\lambda_{11}}{\sqrt{15}}A_2 & \frac{i\lambda_{11}}{\sqrt{10}} A_1 & 0 & 0 & {m_{\Delta}}_{21}-\frac{\lambda_4}{5\sqrt6}A_2
\end{array}
\right),
\end{equation}

\begin{equation}\label{MT21}
B_{21}=\left(
\begin{array}{cccccc}
 B_{H21}&-\frac{i\lambda_{7}}{\sqrt3}A_1 & -\frac{i\sqrt2\lambda_{7}}{3}A_2 & 0 &0 &0 \\
\frac{i\lambda_{10}}{\sqrt3}A_1 & B_{D21} & 0 & -\frac{i\lambda_{11}}{2\sqrt5}A_1 & -\frac{i\lambda_{12}}{2\sqrt5}A_1 & -\frac{i\lambda_{12}}{\sqrt{15}}A_2\\
-\frac{i\sqrt2\lambda_{10}}{3}A_2 & 0  & B'_{D21} &\frac{i\lambda_{11}}{\sqrt{30}} A_2 &-\frac{i\lambda_{12}}{\sqrt{30}} A_2 &-\frac{i\lambda_{12}}{\sqrt{10}} A_1 \\
0 & \frac{i\lambda_{9}}{2\sqrt5}A_1 & -\frac{i\lambda_{9}}{\sqrt{30}} A_2 & {m_{\Delta}}_{21}+\frac{\lambda_4}{5\sqrt6}A_2 & \frac{2\lambda_6}{\sqrt{15}}E & 0\\
0 & \frac{i\lambda_{8}}{2\sqrt5}A_1 & \frac{i\lambda_{8}}{\sqrt{30}} A_2 & \frac{2\lambda_5}{\sqrt{15}}E & {m_{\Delta}}_{12}-\frac{\lambda_3}{5\sqrt6}A_2 & 0\\
0 & \frac{i\lambda_{8}}{\sqrt{15}}A_2 & \frac{i\lambda_{8}}{\sqrt{10}} A_1 & 0 & 0 & {m_{\Delta}}_{12}-\frac{\lambda_3}{5\sqrt6}A_2
\end{array}
\right),
\end{equation}

\begin{equation}\label{MT23}
B_{23}=\left(
\begin{array}{ccc}
0 &0 &0 \\
 -\frac{i\alpha_{3}}{2\sqrt5}A'_1 & -\frac{i\alpha_{4}}{2\sqrt5}A'_1 & -\frac{i\alpha_{4}}{\sqrt{15}}A'_2\\
\frac{i\alpha_{3}}{\sqrt{30}} A'_2 &-\frac{i\alpha_{4}}{\sqrt{30}} A'_2 &-\frac{i\alpha_{4}}{\sqrt{10}} A'_1 \\
 \frac{\alpha_1}{5\sqrt6}A'_2 & 0 & 0\\
 0 & -\frac{\alpha_2}{5\sqrt6}A'_2 & 0\\
 0 & 0 & -\frac{\alpha_2}{5\sqrt6}A'_2
\end{array}
\right),
\end{equation}

\begin{equation}\label{MT32}
B_{32}=\left(
\begin{array}{cccccc}
0 & \frac{i\alpha_{4}}{2\sqrt5}A'_1 & -\frac{i\alpha_{4}}{\sqrt{30}} A'_2 & \frac{\alpha_2}{5\sqrt6}A'_2 & 0 & 0\\
0 & \frac{i\alpha_{3}}{2\sqrt5}A'_1 & \frac{i\alpha_{3}}{\sqrt{30}} A'_2 & 0 & -\frac{\alpha_1}{5\sqrt6}A'_2 & 0\\
0 & \frac{i\alpha_{3}}{\sqrt{15}}A'_2 & \frac{i\alpha_{3}}{\sqrt{10}} A'_1 & 0 & 0 & -\frac{\alpha_1}{5\sqrt6}A'_2
\end{array}
\right),
\end{equation}
and
\begin{equation}\label{MT33}
B_{33}=\left(
\begin{array}{ccc}
\beta_1 S_1 &0 &0 \\
 0& \beta_1 S_1 &0 \\
0& 0 & \beta_1 S_1
\end{array}
\right).
\end{equation}

Here
\begin{eqnarray}
B_{H12}&\equiv &m_H+\frac{i\lambda_1}{\sqrt{6}}A_2+\frac{2\lambda_2}{\sqrt{15}}E, \quad \  \  B_{H21}\equiv  m_H-\frac{i\lambda_{1}}{\sqrt{6}}A_2+\frac{2\lambda_2}{\sqrt{15}}E, \nonumber\\
B_{D12}&\equiv &m_D+\frac{i\lambda_{13}}{3\sqrt{6}}A_2+\frac{4\lambda_{14}}{3\sqrt{15}}E, \quad B_{D21}\equiv m_D-\frac{i\lambda_{13}}{3\sqrt{6}}A_2+\frac{4\lambda_{14}}{3\sqrt{15}}E, \nonumber\\
B'_{D12}&\equiv &m_D+\frac{i\lambda_{13}}{3\sqrt{6}}A_2-\frac{2\lambda_{14}}{\sqrt{15}}E, \quad B'_{D21}\equiv  m_D-\frac{i\lambda_{13}}{3\sqrt{6}}A_2-\frac{2\lambda_{14}}{\sqrt{15}}E.\nonumber
\end{eqnarray}
The mass matrix can be also expressed symbolically as
\begin{equation}\label{trisym1}
M_T=\left(
\begin{array}{ccc}
0_{(6\times 6)} &M_{G(6\times 6)} &0_{(6\times 3)}\\
M_{G(6\times 6)} & 0_{(6\times 6)} &\frac{1}{\sqrt{10}}M_{G(6\times 3)} \\
0_{(3\times 6)} & \frac{1}{\sqrt{10}}M_{G(3\times 6)} & 10M_{G(3\times 3)} \\
\end{array}
\right).
\end{equation}
Note that the texture in (\ref{trisym1}), constrained by the F- and D-flatness conditions,
differs slightly from that in (\ref{m3p3}).
However, as will be seen in the rest of this Section,
the mechanism  of suppressing proton decay following (\ref{m3p3})
will not change.

In SUSY GUTs, the dominant channels inducing proton decay are through the dimension-5 operators \cite{ETM1,ETM2}
\begin{equation}\label{dimension5}
-W_5=C_L^{ijkl}\frac{1}{2}q_iq_jq_kl_l+C_R^{ijkl}u_i^cd_j^cu_k^c e_l^c,
\end{equation}
which are called the $LLLL$ and $RRRR$ operators, respectively,
obtained by integrating out the  color triplet and anti-triplet  Higgs superfields in the interactions in (\ref{superp1}).
Both $C_L^{ijkl}$ and $C_R^{ijkl}$ are inversely proportional to the effective mass of the colored Higgsino.
Since only $B_{11}$ part couples with fermions,
we can get the effective mass by integrating out the uncoupled parts.
From (\ref{trisym1}),  such a mass matrix is similar to the mass matrix  in the double seesaw models for
neutrino masses \cite{double,double2} which is used to generate the small neutrino masses.
In the present  model, the effective masses are large instead of small because $B_{23}\sim B_{32}\ll B_{12}\sim B_{21}\ll B_{33}$.
Similarly, this proton decay suppression mechanism requires two steps of integrations.
Since $S_1$ is ten times of the GUT scale, the $B_{33}$ part can be integrate out first.
Then the mass matrix becomes
\begin{equation}\label{trisym2}
M_T=\left(
\begin{array}{cc}
0_{(6\times 6)} &M_{G(6\times 6)}\\
M_{G(6\times 6)} & M_{I(6\times 6)}
\end{array}
\right),
\end{equation}
where
\begin{equation}\label{seesawmass}
 M_{I(6\times 6)}=-B_{23}\cdot B_{33}^{-1} \cdot B_{32},
\end{equation}
is a matrix with all elements of the order $10^{-2}M_G$.
Then after the second step,
\begin{equation}\label{trisym3}
M_{eff}=-B_{21}\cdot M_{I(6\times 6)}^{-1} \cdot B_{12}\sim\frac{M_{G}^2}{M_I}\sim 2\times 10^{18}\textrm{GeV},
\end{equation}
which is of the order of  the Plank scale, a hundred times heavier than those color triplet Higgs
masses in SUSY GUT models.
The proton decay rates will be suppressed by a factor of $10^{-4}$,
which is small enough surviving all the  current experimental limits.
One may have found that $M_{I(6\times 6)}$ is a rank 3 matrix and is thus not reversible.
This is because that we have not introduced the third \textbf{10}+\textbf{120}-plet $H_3+D_3$ for simplicity.
But, as was discussed in \cite{dlz,dlz2},
each rank contributes one eigenvalue in the effective masses.
The diagonal form of the effective mass matrix is
\begin{equation}\label{effmass}
 M_{eff}=\textrm{diag} \ O(\frac{M_G^2}{M_I}, \frac{M_G^2}{M_I}, \frac{M_G^2}{M_I}, \infty,\infty, \infty).
\end{equation}
Note that it is the lightest eigenvalues that dominates the proton decay rates,
while the three infinitely heavy masses do not contribute.

The suppression can be better understood if we write down the dimension-5 operators explicitly.
The coefficients $C_L$s at the GUT scale $M_G$ are \cite{fukuyamageneral}
\begin{equation}\label{cl}
C_L^{ijkl}(M_G)=\left(
\begin{array}{cccc}
Y^{ij}_{10}, & Y^{ij}_{120}, & Y^{ij}_{120}, & Y^{ij}_{126}
\end{array}
\right)
\left(
\begin{array}{cccc}
(M_T^{-1})_{11} &(M_T^{-1})_{12} & (M_T^{-1})_{13} & (M_T^{-1})_{14}\\
(M_T^{-1})_{21} & (M_T^{-1})_{22}& (M_T^{-1})_{23} & (M_T^{-1})_{24}\\
(M_T^{-1})_{31} & (M_T^{-1})_{32}& (M_T^{-1})_{33} & (M_T^{-1})_{34}\\
(M_T^{-1})_{51} & (M_T^{-1})_{52}& (M_T^{-1})_{53} & (M_T^{-1})_{54}
\end{array}
\right)
\left(
\begin{array}{c}
Y^{kl}_{10}\\
Y^{kl}_{120} \\
Y^{kl}_{120} \\
Y^{kl}_{126}
\end{array}
\right).
\end{equation}
Here the Yukawa couplings are strongly constrained by fitting the fermion masses and mixing \cite{fits1,fits2,fits3,fits4}. The elements of $M_T^{-1}$ are of the order $\frac{1}{M_G}$ in usual SUSY GUT models, but in our model, %the symbolical matrix of $M_T^{-1}$ is
\begin{equation}\label{triinv}
M_T^{-1}=\frac{1}{M_G}\left(
\begin{array}{ccc}
10^{-2}_{(6\times 6)} &1_{(6\times 6)} &10^{-\frac{3}{2}}_{(6\times 3)}\\
1_{(6\times 6)} & 0_{(6\times 6)} &0_{(6\times 3)} \\
10^{-\frac{3}{2}}_{(3\times 6)} & 0_{(3\times 6)} & \frac{1}{10}_{(3\times 3)} \\
\end{array}
\right).
\end{equation}
We can see clearly that the elements contributing to dimension-5 operators, i.e. elements in the up-left most block, are of the order $\frac{10^{-2}}{M_G}$. This is the same conclusion  drawn in (\ref{trisym3}).
This conclusion applies for both the LLLL and RRRR operators.

\section{The weak doublets} \label{DT}

Like in \cite{dlz2}, the doublet-triplet splitting (DTS) problem requires a minimal fine-tuning, and similar results can be reached. The up-type doublets are ordered as
\begin{equation}%\label{phiu}
\varphi_{u}=(H_{1u}, D_{1u}, D'_{1u}, \Delta_{1u}, \overline{\Delta}_{1u}; H_{2u}, D_{2u}, D'_{2u}, \Delta_{2u}, \overline{\Delta}_{2u}; \Delta_{3u}, \overline{\Delta}_{3u}),
\end{equation}
while down-type doublets are
\begin{equation}%\label{phid}
\varphi_{d}=(H_{1d}, D_{1d}, D'_{1d}, \overline{\Delta}_{1d}, \Delta_{1d}; H_{2d}, D_{2d}, D'_{2d}, \overline{\Delta}_{2d}, \Delta_{2d}; \overline{\Delta}_{3d}, \Delta_{3d}).
\end{equation}
The mass terms of the weak doublets are  given by $(\varphi_{d})_a(M_D)_{ab}(\varphi_u)_b$, with the $12\times 12$ matrix $M_D$ written as
\begin{equation}%\label{doublet}
M_D=\left(
\begin{array}{ccc}
0_{(5\times 5)} &A_{12(5\times 5)} &0_{(5\times 2)}\\
A_{21(5\times 5)} & 0_{(5\times 5)} &A_{23(5\times 2)} \\
0_{(2\times 5)} & A_{32(2\times 5)} &A_{33(2\times 2)}
\end{array}
\right)=\left(
\begin{array}{ccc}
0_{(5\times 5)} &M_{G(5\times 5)} &0_{(5\times 2)}\\
M_{G(5\times 5)} & 0_{(5\times 5)} &\frac{1}{\sqrt{10}}M_{G(5\times 2)} \\
0_{(2\times 5)} & \frac{1}{\sqrt{10}}M_{G(2\times 5)} & 10M_{G(2\times 2)}
\end{array}
\right).
\end{equation}

For general parameters, there is no massless doublet.
The DTS, which requires a zero determinant of $M_D$, can be obtained if
\begin{equation}\label{detmd}
\textrm{Det}(M_D)=\textrm{Det}(A_{12})*\textrm{Det}(A_{21})*\textrm{Det}(A_{33})=0.
\end{equation}
$\textrm{Det}(A_{33})$ is obviously nonzero, leaving us two choices. $\textrm{Det}(A_{12})=0$
is not acceptable because the large top quark mass would not be generated for perturbative Yukawa couplings.
If we chose $\textrm{Det}(A_{21})=0$, we will further get the massless doublet can be expressed as
\begin{equation}\label{phiud}
 H_{u} =\sum_{i=1}^{13} \alpha_u^{i\ast} \varphi_{u}^i,\quad  H_{d} =\sum_{i=1}^{13} \alpha_d^{i\ast} \varphi_{d}^i.
\end{equation}
and the components are, up to a normalization factor,
\begin{eqnarray}
\alpha_u^\ast&=&O(\underbrace{1,...,1}_\textrm{five};\underbrace{0,...,0}_\textrm{seven}),\label{phiu}\nonumber\\
\alpha_d^\ast&=&O(\underbrace{10^{-2},...,10^{-2}}_\textrm{five}; \underbrace{1,...,1}_\textrm{five}; \underbrace{10^{-\frac{3}{2}},...,10^{-\frac{3}{2}}}_\textrm{two}).
\end{eqnarray}
The large ratio of $\frac{\alpha_u^i}{\alpha_d^i}$ $(i\le 5)$ is consistent
with the ratio of $\frac{m_t}{m_b}\sim 100$ at high energy \cite{fits1,fits2,fits3,fits4}.
It also gives the constrain on $\textrm{tan}\beta$
\begin{equation}\label{mtmb}
\textrm{tan}\beta =\frac{v_u}{v_d} \approx \frac{m_t}{m_b}10^{-2}\sim O(1).
\end{equation}
Equation (\ref{mtmb}) suggests that a small $\textrm{tan}\beta$ is favored in the present model,
which is also the same conclusion drawn in \cite{dlz2}.

\section{Summery and conclusions} \label{Sum}

As in \cite{dlz2}, we do not perform explicitly the fine-tuning in the weak doublets which takes only one
free parameter in the superpotential. As was pointed out in \cite{Bajc}, threshold effects can be big in the minimal SUSY SU(5) theory, and can be even bigger due to the more super-heavy particles in SO(10) models. In this work, we have focused mainly on the new method of proton decay suppression and this method does not require explicit threshold effect calculations. The other reason is that there are more than enough free parameters in the superpotential that can be adjusted in calculating the threshold effects to fulfill the gauge coupling unification.

In the present work we have presented a renormalizable SUSY SO(10) model with
sufficient suppression of proton decay.
Similar to  \cite{dlz2},
gauge coupling unification is maintained due to the absence of intermediate scales,
and the seesaw VEV, proton decay and $\textrm{tan}\beta$ are found to be all related,
Thus the main conclusions are quite general in a class of models which follow the mechanisms
of suppressing proton decay through constructing seesaw-like textures in the color triplet mass matrices.
Different from the previous study,
we use \textbf{45}+\textbf{54} instead of \textbf{210} to break SO(10).
Instead of a global U(1) used in \cite{dlz2}, we use
an anomalous U(1) to generate the seesaw VEV through Green-Schwarz mechanism.
We have also included \textbf{120}-plet Higgs to couple with fermions so
that the model is highly realistic.
We have, however, two main problems untouched.
The first is the DTS problem which we simply use an assumed fine-tuning in the weak doublets.
The second is the perturbative difficulty for the gauge coupling above the GUT scale which is also common to all realistic SUSY GUT models.

We can compare our work with \cite{Aulakh1,Aulakh2} where
Higgs in $\textbf{10}+\textbf{120}+\textbf{126}/\overline{\textbf{126}}$ are used to fit fermion masses and mixing
while \textbf{210} is used to break SO(10).
Proton decay suppression is carried out by raising the GUT scale up to the Planck scale or even higher
so that the color-triplet Higgs masses are also enhanced accordingly,
otherwise proton lifetime is around $10^{28}yr$ only.
This picture conflicts with the most important results supporting SUSY GUT which suggest the GUT scale to
be $M_G\sim 2\times 10^{16}$GeV \cite{MSSM1,MSSM2,MSSM3,MSSM4}.
In our work, the suppression of proton decay is realized by enhancement of the effective triplet masses
and the unification scale remains at $M_G$.

%%%%%%%%%%%%%%%%%%%%%%%%%%%%%%%%%%%%%%%%%%%%%%%%%%%%%%%%%%%%%%%%%%%%%%%%%%%%%%%
% Bibliography
%%%%%%%%%%%%%%%%%%%%%%%%%%%%%%%%%%%%%%%%%%%%%%%%%%%%%%%%%%%%%%%%%%%%%%%%%%%%%%%

\end{document}